\documentclass[aps,prl,twocolumn,showpacs,superscriptaddress,groupedaddress]{revtex4-1}  
\usepackage{dcolumn}   
\usepackage{bm}        
\usepackage{amssymb}   
 \usepackage{amsmath}

\usepackage{natbib}
\usepackage{float}
\usepackage[dvips,final]{graphicx}
\usepackage{pslatex}
\usepackage{relsize}
\usepackage{bm}
\usepackage{xspace} 

\usepackage{xcolor}
\usepackage{soul}

\def\ba{\begin{eqnarray}}
\def\ea{\end{eqnarray}}
\def\be{\begin{equation}}
\def\ee{\end{equation}}

\bibpunct{[}{]}{,}{n}{}{} 

\begin{document}
\title{Typical skyrmions versus bimerons: a long-distance competition in ferromagnetic racetracks}

\author{A. S. Ara\'{u}jo}
\affiliation{Universidade Federal de Vi\c cosa, Departamento de F\'isica,
Avenida Peter Henry Rolfs s/n, 36570-000, Vi\c cosa, MG, Brazil}
\author{R. L. Silva}
\author{R. C. Silva}
\affiliation{Departamento de Ci\^encias Naturais, Universidade Federal do Esp\'irito Santo, S\~ao Mateus, ES, 29932-540, Brazil.}
\author{R. J. C. Lopes}
\affiliation{Universidade Federal de Vi\c cosa, Departamento de F\'isica,
Avenida Peter Henry Rolfs s/n, 36570-000, Vi\c cosa, MG, Brazil}
\author{D. Altbir}
\affiliation{Departamento de F\'isica, CEDENNA, Universidad de Santiago de Chile, USACH, Av. Ecuador 3493, Santiago, Chile}
\author{V. L. Carvalho-Santos}
\author{A. R. Pereira}
\affiliation{Universidade Federal de Vi\c cosa, Departamento de F\'isica,
Avenida Peter Henry Rolfs s/n, 36570-000, Vi\c cosa, MG, Brazil}


\begin{abstract}
 During the last years, topologically protected collective modes of the magnetization have called much attention. Among these, skyrmions and merons have been the object of intense study. In particular, topological skyrmions are objects with an integer skyrmion number $Q$ while merons have a half-integer skyrmion charge $q$. In this work, we consider a $Q=1$ skyrmion, composed by a meron and an antimeron (bimeron), displacing in a ferromagnetic racetrack, disputing a long-distance competition with its more famous counterpart, the typical $Q=1$ cylindrically symmetrical skyrmion. Both types of topological structures induce a Magnus force and then are subject to the Hall effect. The influence of the Dzyaloshinskii-Moriya interaction ($DMI$) present in certain materials and able to induces $DMI$-skyrmions is also analyzed. Our main aim is to compare the motions (induced by a spin-polarized current) of these objects along with their own specific racetracks. We also investigate some favorable factors which are able to give breath to the competitors, impelling them to remain in the race for longer distances before their annihilation at the racetrack lateral border. An interesting result is that the $DMI$-skyrmion  loses this hypothetical race due to its larger rigidity.
 \end{abstract}

\flushbottom \maketitle
%
%
\thispagestyle{empty}

\section*{Introduction}

Skyrmions \cite{Skyrme} are topologically protected states that have been introduced in the framework of the two-dimensional ($2d$) Heisenberg model (HM) by Belavin and Polyakov \cite{BP}. The ($2d$) HM is defined by the Hamiltonian $H = -J \sum_{\{i,j\}}\vec{S}_{i}\cdot \vec{S}_{j}$,
where $J>0$ is the ferromagnetic coupling constant, the sum is over nearest-neighbor spins and the spin field $\vec{S}(\vec{x})$ obeys the constraint
$\vec{S}^{2}(\vec{x})=S_{x}^{2}(\vec{x})+S_{y}^{2}(\vec{x})+S_{z}^{2}(\vec{x})=S^{2}$, with $S$ being a constant. Topologically, skyrmions correspond to the mapping of the spin-space sphere $(\sum^{int})_{2}$ onto the continuum plane $\vec{r}=(x,y)$ (physical space $(\sum^{phy})_{2}$). Consequently,
they are characterized by a skyrmion integer number $Q= \pm 1, \pm2, ...$, and have finite energy $E_{s}=4\pi JS^{2}\mid Q \mid$, independent of the skyrmion size $R$  since the continuum limit of the Heisenberg model is scale-invariant.

Considering the mapping $(\sum^{int})_{2} \rightarrow (\sum^{phy})_{2}$, the Belavin-Polyakov skyrmion configurations can have essentially two faces as seen by different perspectives, which depend on the boundary conditions (or stereographic projection). For $\vec{S}(\vec{r})\rightarrow(0,0,\pm S)$ as $\vec{r} \rightarrow \infty$, one gets the $\mid Q \mid$ core configuration (type $I$-skyrmion) while for $\vec{S}(\vec{r})\rightarrow(\pm S,0,0)$ as $r \rightarrow \infty$, one gets the $2\mid Q \mid $ core configuration (type $II$-skyrmion). For the same $Q$, both skyrmions (type $I$ and type $II$) have the same energy. Therefore, we mean that the core occupies a small localized region in which $S_{x}^{2}+S_{y}^{2} = 0$ and consequently, $S_{z}=\pm S$. However, depending on parameters like small anisotropies, external magnetic fields, or others that should favor out-of-plane or in-plane spins, structures similar to type $I$ or type $II$ skyrmions, respectively, could be excited in a system.

In our study we consider $Q=\pm 1$ skyrmions since they are energetically favorable. Because type $I$-skyrmions exhibit great potential to be used in storage and processing-information technologies, much attention has been dedicated to study such a spin texture \cite{Yu2010,Heinse2011}. However, for those applications, some intrinsic difficulties in generating and guiding them along a nanostripe need to be overcome. For instance, to use them in spintronic applications, the main barrier is the inability to move skyrmions straight along applied currents. Indeed, it is well known that type $I$-skyrmions suffer the effect of the Magnus force, which leads to the skyrmion Hall effect. Amongst some theoretical propositions to suppress the skyrmion Hall effect, there are possibilities of engineering magnetic materials \cite{Toscano}, the formation of coupled skyrmions displacing in bilayer compounds \cite{R1,R2,R3} and spin-current driven skyrmion dynamics \cite{R4}. Based on the above, it should be relevant to see what occurs with the dynamics of topological structures with different shapes along pathways to get more insights to prevail over intrinsic technological difficulties.

In this paper we  give attention to type $II$-skyrmion  textures also called bimerons \cite{Zhang2020}. These objects are not cylindrically symmetric\cite{Tretiakov07,MarcoAnto2009} and may have also important consequences in quantum magnetism. For instance, considering $2d$-antiferromagnets with general spin $S$ and the case $Q=1$, the merons\cite{Gross78} forming a double core skyrmion \cite{Fernandes2019} are ``spin-$S$ spinons" \cite{Baskaran03, Antonio07}, which appear as essential objects in the seek for two-dimensional quantum spin liquid\cite{Anderson87} states of spin-half ($S=1/2$). On the other hand, another kind of bimeron structures may also be found in thin
chiral magnetic films\cite{Ezawa2011} induced by nonmagnetic impurities\cite{Ricardo2014}, as well as stabilized in confined geometries\cite{Iakovlev2018}.

The main goal of this paper is to analyze the trajectories of  both types of skyrmions described above in ferromagnetic racetracks. In principle, it is shown that if we consider a massless model to describe the dynamics of a bimeron, its trajectory and velocity along a nanotrack would be the same as that predicted for type $I$-skyrmions. Nevertheless, due to its non-cylindrical symmetry, the displacement of bimerons mass-center induces an effective mass which is different from the mass of its type $I$-skyrmion counterpart. Therefore, it should move in a straight line for longer/shorter distances. Thus, by means of analytical calculations and micromagnetic simulations, we study type $II$-skyrmions focusing on  their sensitivity to the Magnus force. The results are compared with the trajectories obtained for type $I$-skyrmions. Here we have to distinguish two categories of type $I$-skyrmions, which depends on the specific materials they can reside: type $I$-skyrmions living in ferromagnetic materials with Dzyaloshinskii-Moriya interaction, described by a coupling constant $D$, added to the Heisenberg Hamiltonian $H$ and genuine type I-skyrmions which subsist in ferromagnets without Dzyaloshinskii-Moriya interaction ($DMI$). Although they have very similar shapes, the small and basic contrasts between them may lead to different dynamics. For instance, when the $DMI$ is present, the skyrmion has a more rigid structure and its size (controlled by the ratio $D/J$) remains practically constant during movement. For racetrack materials with $DMI$, hereafter, the skyrmions will be called $DMI$-skyrmions while the name $I$-skyrmions, will be held for the natural counterpart of type $II$-skyrmions.

\section*{Theoretical model}

Type $II$-skyrmions or bimerons have two centers in which a meron and an antimeron are positioned. A meron with a winding number $\eta=\pm 1$ and core polarization $P=\pm 1$ has a half-integer topological charge $q=\eta P/2$ (the meron wraps only half of the sphere). Therefore, a pair constituted by a meron ($\eta=1$) and an antimeron ($\eta=-1$) with the same polarization (for example $P=1$) has opposite skyrmion numbers adding to zero ($Q=0$) and thus, such a pair belongs to the same topological sector as uniform ground states. This object would be then topologically unstable since it can be deformed continuously into a ground state with zero skyrmion number. On the other hand, if a pair has a meron and an antimeron with antiparallel core polarizations, these half-integer structures would have equal skyrmion numbers adding to a total of $+1$ or $-1$, belonging to a nontrivial topological sector and thus cannot be deformed continuously into a ground state. It is exactly what occurs with bimerons, which are characterized by a topological invariant (the skyrmion number), defined as
\begin{eqnarray}\label{ChargeQ}
Q = \frac{1}{8\pi}\int d^{2}\vec{x}\epsilon_{ij} \epsilon_{\alpha
\beta \delta} n_{\alpha}\partial_{i}n_{\beta}
\partial_{j}n_{\gamma},
\end{eqnarray}\\
where $\hat{n}(\vec{x})=\vec{S}/S$ is the unit vector parallel to the local magnetization $\vec{S}(\vec{x})$.

\begin{figure}[hbt]
\begin{center}
\includegraphics[width=80.0mm]{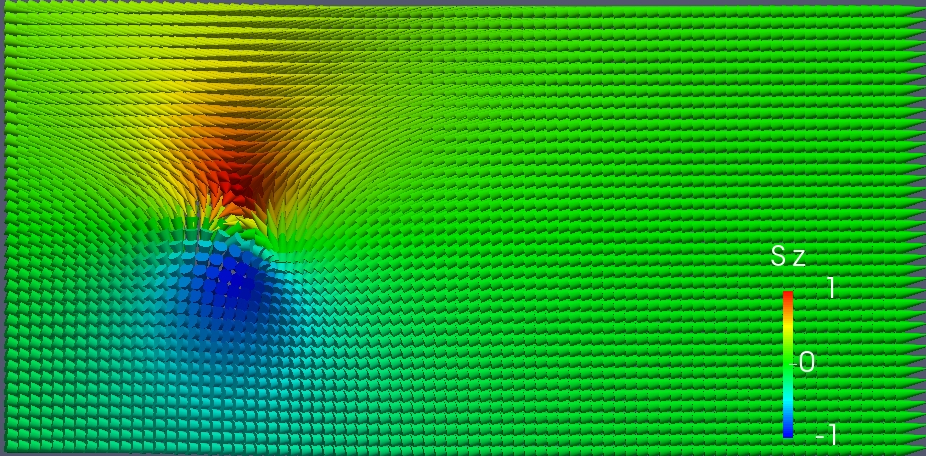}
\includegraphics[width=80.0mm]{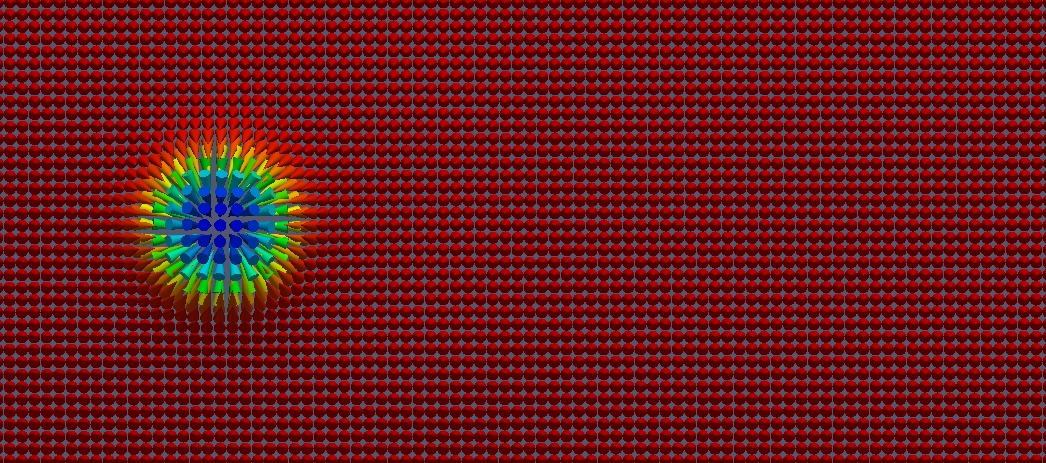}
\caption{\label{fig2}  Spin projection along the $z$-axis (normal to the racetrack plane) is depicted in color. $a)$ depict a bimeron (type $II$-skyrmion) in a vertical position ($v$-bimeron). $b)$ representation of a type $I$-skyrmion texture. Each type of skyrmion runs in its own lane. Their coexistence in the same material is not a trivial possibility since they live in systems with different tendencies for spin arrangements (in-plane or out-of-plane).}
\end{center}
\end{figure}

The continuum limit of the $2d$-isotropic ferromagnet described by a Hamiltonian $H$ consists in the famous nonlinear $\sigma$-model, given by $(J/2) \int d^{2}\vec{x}(\partial_{\nu}\vec{S})^{2}$, $\nu=1,2$ and the constraint $S^{2}=1$ (without loss of generality, we use an unit
spin vector). The explicit static spin configuration of a bimeron can be obtained by using boundary conditions $\vec{S}\rightarrow (1,0,0)$ at $\vec{r}\rightarrow \infty$. Then, parametrizing the spin vector $\vec{S}(\vec{r})$ by two scalar fields, the polar and azimuthal
angles $\theta$ and $\phi$, $\vec{S}=(\cos\theta \cos\phi,\sin\theta \sin\phi, \cos\theta)$, this static solution with $Q=1$
(energy equal to $4\pi J$), size $R$ (merons separated by a distance $R$) and mass center localized at the origin can be written as

\begin{subequations}\label{thetaEq}
\begin{equation} \label{twocore}
\theta_{2c}^{^v(h)}=\arccos \left(\frac{R\,c_{i}}{\rho^2+R^{2}/4}\right),
\end{equation}
\begin{equation} \label{phiEq}
\phi_{2c}^{v(h)}=\arctan\left(\frac{c_{i}-R/2}{c_j}\right)-\arctan\left(\frac{c_i+R/2}{c_j}\right)\,,
\end{equation}
\end{subequations}

\noindent
where $\rho=\sqrt{\zeta x^2+\xi y^2}$, and $(c_i,c_j)=(x,y)$ and $(c_i,c_j)=(y,x)$ for type $II$-skyrmions with the cores aligned horizontally ($h$-bimeron) and vertically ($v$-bimeron), respectively. If $\zeta=\xi$, we obtain a regular rigid bimeron in which the two cores are not deformed. If $\zeta\neq \xi$, we obtain a bimeron having an elliptical shape. In Fig. \ref{fig2}-a, we show the vector field of the above described model for a $v$-bimeron with the cores aligned vertically, and $\zeta=\xi=1$.

Aiming to compare the dynamics of $I$- and $II$-skyrmion, we can describe a $Q=1$ $I$-skyrmion solution (Fig. \ref{fig2}-b) with characteristic radius $R$, energy equal to $4\pi J$, and placed at $(0,0)$, as
\begin{equation} \label{twocore2}
\theta_{1c}=\arccos \left(\frac{R^{2}-\rho^2}{R^2+\rho^2}\right), \hspace{0.5cm}
\phi_{1c}=\arctan\left(\frac{y}{x}\right)\,.
\end{equation}

.

Micromagnetic simulations are performed to study the stabilization and dynamics of these skyrmion structures. Firstly, we have stabilized the bimeron in a racetrack composed by an isotropic Heisenberg ferromagnetic material (Fig.\ref{fig2}-a) at zero temperature by relaxation and using the solutions of the $O(3)$ nonlinear $\sigma$-model given by expressions \ref{thetaEq}. The investigated racetrack has a width (distance between the upper and lower lateral borders) equal to $L_{y}=80a$ and length $L_{x}=300a$, where $a$ is the lattice parameter. The calculations consider periodic boundary conditions along the $x$-direction and open boundary condition along the $y$-direction. The bimeron was stabilized with the following parameters: $J=1$ and $R= 4a$. Similar parameters are also used for type $I$-skyrmion (Fig.\ref{fig2}-b). Both tracks in Fig.\ref{fig2} are organized in parallel to simulate a hypothetical race between the $I$- and $II$-skyrmions. Since we are studying $4$ structures ($I$-skyrmion, $DMI$-skyrmion, $h$-bimeron and $v$-bimeron), our imaginary running track is constituted by $4$ lanes, each one made by a ferromagnetic material with characteristics able to support its resident competitor.

After having stabilized the bimeron, adjusting its configuration inside the given racetrack, fourth-order Runge-Kutta method is employed to
compute the dynamics of the magnetic moment, $\vec{S}_{i}$, by solving the the Landau-Lifshitz-Gilbert ($LLG$) equation\cite{Landau,Gilbert},
\begin{equation}\label{eq2}
\frac{\partial\vec{S}_{i}}{\partial t} = -\gamma\vec{S}_{i}\times
\hat{H}_{\text{eff}}^{i}+\alpha \vec{S}_{i}\times\frac{
\partial\vec{S}_{i}}{\partial t}
\end{equation}

\noindent
where $\gamma$ is the gyromagnetic ratio, $\hat{H}_{\text{eff}}^{i}=-\frac{1}{\mu_{s}}\frac{\partial\mathcal{H}}{\partial \vec{S}_{i}}$ is the net effective magnetic field on each spin, and $\alpha$ is the Gilbert damping coefficient. The spin-polarized current is
introduced by using the Berger spin-transfer torque\cite{Berger}:
\begin{equation}\label{eq3}
\vec{\tau}_{B}= p\left(\vec{j}\cdot\nabla\right)\vec{S }\,,
\end{equation}
and
\begin{equation}\label{eq4}
\vec{\tau}_{B\beta}= p\beta\vec{S}\times\left(\vec
{j}\cdot\nabla\right)\vec{S}\,,
\end{equation}

\noindent
where Eq.(\ref{eq3}) and Eq.(\ref{eq4}) are the adiabatic and non-adiabatic torque, respectively. Here $p$ is the spin polarization of the electric current density $\vec{j}$, while $\beta$-parameter characterizes its relative strength to the Berger's torque (Eq.(\ref{eq3})).

\begin{figure}
\includegraphics[scale=0.13]{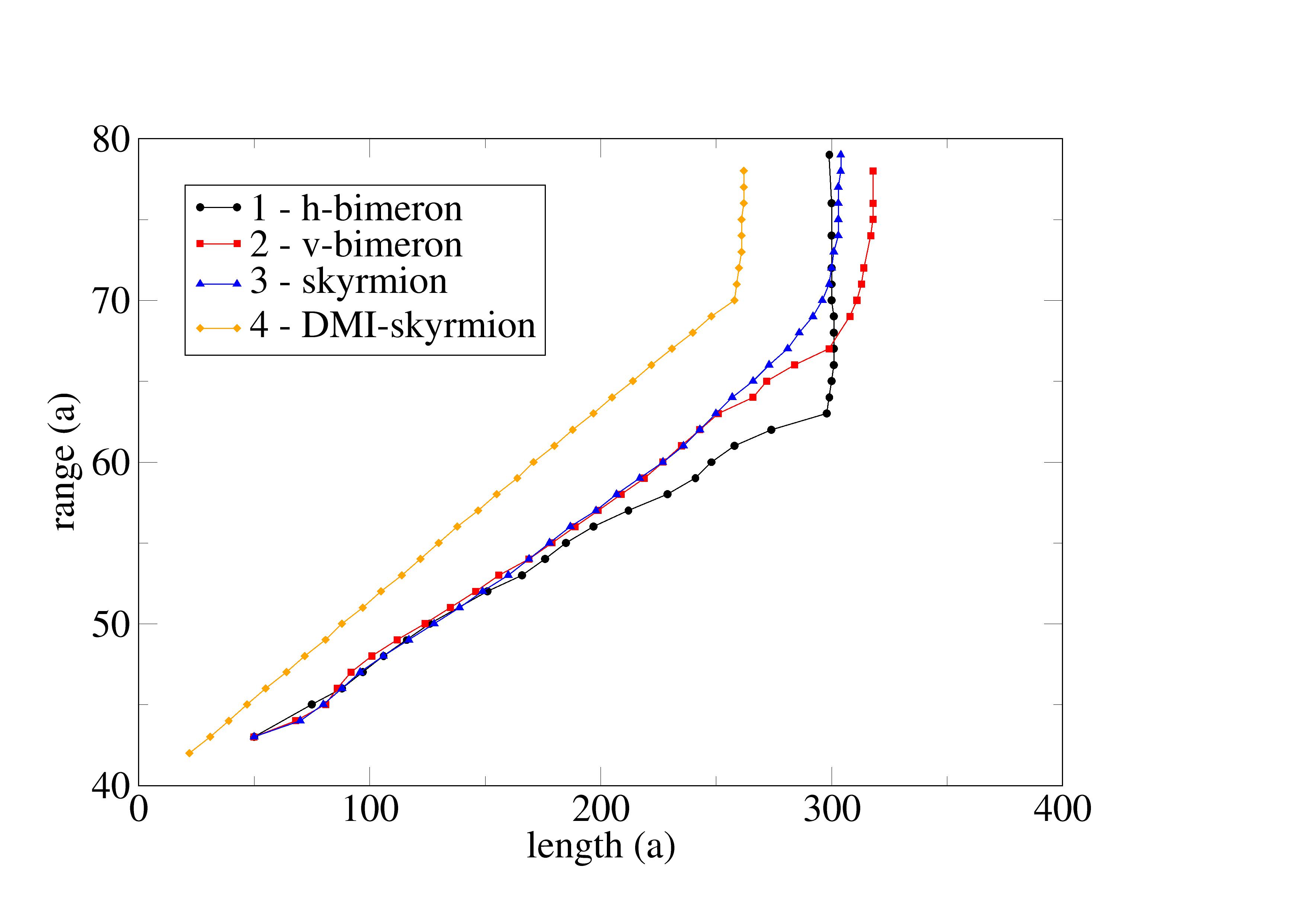}\caption{Trajectories described by the skyrmions (all with $R=4a$ during their motions in a racetrack with width $L_{y}=80a$ and length $L_{x}=300a$. Black, red, and blue lines depict the trajectories of the $h$-bimeron, $v$-bimeron, and $I$-skyrmion, respectively. Orange line depicts the trajectory of $DMI$-skyrmion. In a hypothetical race among these objects, the $h$-bimeron would be the winner.}
\label{Trajectories}
\end{figure}

\section*{Results}

\begin{figure}
\includegraphics[scale=0.3]{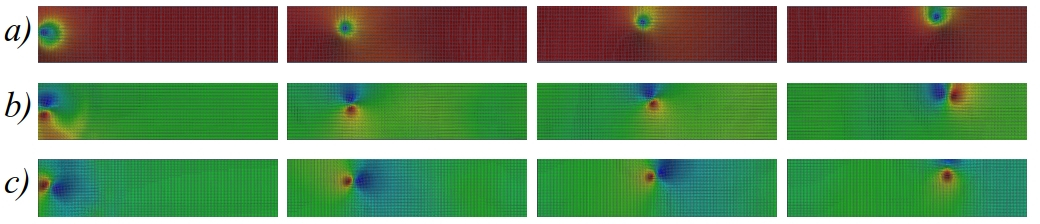}\caption{Snapshots for $4$ subsequent times of the investigated skyrmions in their appropriated tracks during a hypothetical race. Here, it is shown only racetracks made of ferromagnetic materials without $DMI$; $a$, $b$, and $c$ present the evolution of the dynamics of $I$-skyrmion, $v$-bimeron, and $h$-bimeron, respectively. }
\label{Snapshots}
\end{figure}

After stabilizing the skyrmions, we have performed micromagnetic simulations to obtain their mass center position as a function of time for four configurations: \textit{i}) a $DMI$-skyrmion; \textit{ii}) a $I$-skyrmion; \textit{iii}) a $v$-bimeron; and \textit{iv}) a $h$-bimeron. Before presenting the main results, we have to say something about the particularities of the above structures. Specifically, different from $I$- and $II$-skyrmions, $DMI$-skyrmions demand extra parameters and factors to be stabilized in a magnetic compound, such as the coupling $D$ and the presence of an external magnetic field along the direction perpendicular to the magnetic plane. Instead of using the field, we stabilize this kind of structure by a small easy-axis anisotropy $k_{z}/J=0.11$. In addition, we use $D/J=0.26$ for the Dzyaloshinskii-Moriya coupling constant. These factors convert $DMI$-skyrmion configurations in rigid structures, much more inflexible than the other skyrmions investigated here. Indeed, $DMI$-structures are heavier than the other skyrmions and their size does not suffer significant variation during their motions as will be discussed below. This hardness is not expected for $I$- and $II$-skyrmions, since they are described only by a Heisenberg Hamiltonian. As a consequence, their sizes may suffer some fluctuations during their motion, mainly when the spin current is initially applied. Further, at first sight, because the $II$-skyrmion has two merons with opposite winding numbers, one may expect that the
meron tends to suffer the Magnus force impelling it to, let's say, the upper border, while its counterpart antimeron tends to go to the opposite side, i.e., the lower border (see Fig.\ref{fig2}-a). Nevertheless, the type $II$-skyrmion as a whole has a topological number $Q=1$ and, therefore, it tends to suffer the Magnus force, similar to what happens to $I$-skyrmion (all that depends on $q=\eta P/2$; both merons of the bimeron have positive charge $q=1/2$, moving in the same direction). In other words, the total Magnus force on the structure as a whole is not zero. Therefore, the bimeron mass center moves along the racetrack suffering the skyrmion Hall effect. The obtained results here confirm this statement. In Fig. \ref{Trajectories} we present the respective trajectories followed by the four types of structures during their motions. Firstly, we notice that the deviation from a straight  trajectory of a $DMI$-skyrmion (orange line) is greater than all the other ones. That is, if the $DMI$-skyrmion center starts its motion at  the same point of the other structures, it reaches the y-border at a smaller position along the $x$-axis. Additionally, it can be observed that $h$-bimerons suffer a smaller deviation due to the skyrmion Hall effect. Indeed, considering the three skyrmions in materials without $DMI$, it can be observed that, until the position $x=250\,a$, the $h$-bimeron occupies a lower position in the $y$-axis when compared to the $v$-bimeron and $I$-skyrmion. Additionally, the trajectories of  $II$-skyrmions are longer than that of the usual type $I$-skyrmions. On the other hand, since $II$-skyrmions contain two centers, their movements must not occur keeping a rigid structure, mainly because the racetrack has a finite size. Indeed, the skyrmion may rotate slightly around its mass center and the two merons could have small vibrations during this process. This makes the $II$-skyrmions  displace faster along the $y$-direction when they are near the border of the stripe and they are annihilated almost at the same time as the $I$-skyrmion is at the track border (See Fig. \ref{Snapshots} and the movies available as supplemental materials online).

To understand the above described results, we will make use of an analytical model, assuming that skyrmions are rigid structures. This assumption is suitable for $DMI$-skyrmions and applies only in a first approximation for the Belavin-Polyakov configurations also treated here. Indeed, for an infinite system, if only exchange interaction is considered, the energy of the $II$-skyrmions is independent of the distance $R$ between the meron and the antimeron centers. The same is valid for type $I$-skyrmion. In this context, the dynamical description of the merons motion can be given by an analytical model neglecting dynamical deformations of the $II$-skyrmion in such a way that the $LLG$ equation can be reduced to the Thiele equation\cite{Thiele}, written as

\begin{eqnarray}
\mathcal{M} \dot{{\bf v}}(t)
+ g \hat{z} \times ({\bf v}(t)-\mathbf{v}_s)
+ {\bf \mathcal{D}} (\alpha {\bf v}(t)-\beta\mathbf{v}_s)
= {\bf F}\,,
\label{Thiele}
\end{eqnarray}
where the first contribution consists of an analogous to Newton's second law, with $\mathcal{M}$ being the effective mass of the collective mode of magnetization, where the mass matrix is given by

\begin{equation}
 \mathcal{M}^{ij} = \frac{1}{\alpha\gamma^{2}}\int \mathrm{d}^2x \left( \partial^{i}\vec{n} \cdot \partial^{j}\vec{n} \right)\,.
\end{equation}

\noindent
The second term in the Thiele equation describes the Magnus force exerted by the magnetic texture in the collective mode of the magnetization, which displaces with velocity $\mathbf{v}_j$ under the action of the spin current, whose spin velocity parallel to the
spin current is $\mathbf{v}_s$. The third contribution in Eq. (\ref{Thiele}) consists of a dissipative force, with $\mathcal{D}$ being the dissipative dyadic, given by $D^{ij}=\alpha\gamma^{2}\mathcal{M}^{ij}$. If we consider the parametrization described by Eqs. (\ref{thetaEq}) and (\ref{twocore2}), with $\zeta=\xi=1$, the effective mass of $I$- and $II$-skyrmions are the same, given by $ \mathcal{M}^{11}_s =  \mathcal{M}^{22}_s \equiv \mathcal{M}_s=8\pi\,b\,(\alpha\gamma^{2}\sqrt{R^2+4\,b^2}\,)^{-1}$ and $ \mathcal{M}^{12}_s =  \mathcal{M}^{21}_s=0$, where $2b=L_{y}$ is the width of the track and we have considered that $L_{y}\gg R$. Under these assumptions, the spatial coordinates of all skyrmion structures are obtained from the solution of Eq. (\ref{Thiele}), evaluated as

\begin{equation} \label{xt}
x(t)=\frac{g^2\,v_s}{{g^2+\alpha^2\mathcal{D}_s^2}}\,t\,,\hspace{0.5cm}y(t)=\frac{g\,\mathcal{D}_s\,v_s}{g^2+\alpha^2\mathcal{D}_s^2}\,\alpha\,t\,.
\end{equation}

\noindent
After eliminating the parameter $t$ (time), we get the trajectory equation $y(x)= (\mathcal{D}_{s} \alpha /g) x$. Note that the function $y(x)$ has a linear dependence on $x$-variable with inclination $\Delta = \mathcal{D}_{s} \alpha /g \propto \mathcal{M}_s $. The trajectory equation $y(x)$ can be directly compared with the simulation results of Fig.\ref{Trajectories}. Indeed, this figure shows that the trajectory of all skyrmions obeys an approximated linear dependence $y_{i}(x)=k_{i}x$ (here, $i=1,2,3,4$ refers to the different types of skyrmions). However, the linear behavior of the simulation results prevails only up to a certain critical value of the $x$-coordinate (let's say, $x_{i,c}$). This is a critical position for the skyrmion in a racetrack, marking the point where the interaction skyrmion-border becomes strong enough to deform the skyrmion configuration, invalidating the application of our analytical results (extra forces should be considered in Eq. (\ref{Thiele})). Point $(y(x_{i,c}), x_{i,c})$ denotes the position in which the skyrmion $i$ finds its ultimate moments. After $(y(x_{i,c}), x_{i,c})$ , the simulations show that the coordinate $y(t)$ increases rapidly with $t$ while $x(t)$ becomes essentially constant ($x(t) \sim x_{i,c})$ (see again Fig.\ref{Trajectories}).

The analytical trajectory equation obtained above explains the accentuated difference between the trajectories of skyrmions in materials with and without $DMI$, as seen in Fig. \ref{Trajectories}. Since the presence of the Dzyaloshinskii-Moriya interaction diminishes the skyrmion radius, the effective mass of the $DMI$-skyrmion is greater than that of the other structures considered here. The initial impact of the spin current on $I$- and $II$-skyrmions increases their sizes appreciably as observed in the simulations, and snapshots of Fig. \ref{Snapshots} can give a clear idea about this behavior. Therefore, remembering that $\Delta \propto \mathcal{M}_s $, then  the deviation of the $DMI$-skyrmion trajectory is larger than that of all other skyrmions residing in materials without $DMI$. Consequently, $DMI$-skyrmion reaches a lower position along the $x$-axis, having a smaller critical $x$-position, confirming  simulation results. On the other hand, our analytical calculations imply that all other skyrmions analyzed here, residing in materials without Dzyaloshinskii-Moriya interaction, have equal masses and consequently they should follow similar trajectories. A  comparison with simulations of Fig. \ref{Trajectories} shows that it is true only in certain \text{parts} of the skyrmion routes (around half-way, $x \sim 150a$). After that, the $h$-skyrmion, $v$-skyrmion and $I$-skyrmion mass-center trajectories disjoint and each skyrmion follows different ways. As a result, their annihilations occur in slightly different $x$-positions.
\begin{figure}
\includegraphics[scale=0.32]{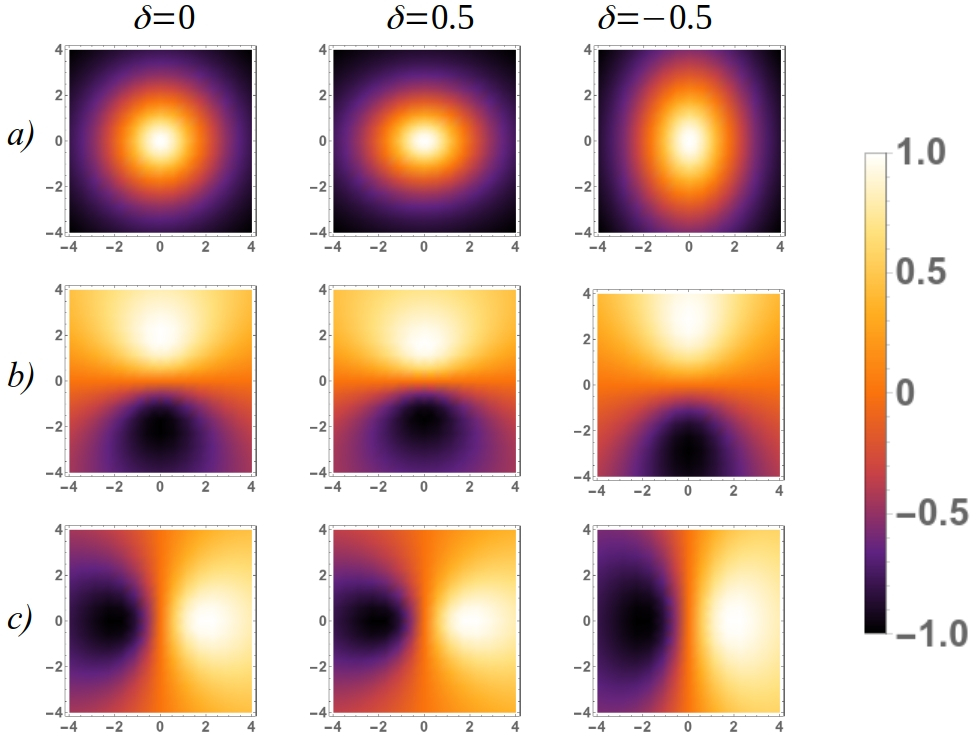}\caption{Density plot of the $m_z$ component of the magnetization of the considered configurations. $a$, $b$, and $c$ show respectively the skyrmion, $v$-meron, and $h$-meron for different values of $\delta$.}\label{Sk-deformados}
\end{figure}

Trying to explain the small differences that occur in the trajectories of type $I$-skyrmion and the bimerons (even the initial position of the bimeron affects its route, causing differences in the trajectories of $v$- and $h$-bimerons), we will assume that there are small deformations in the skyrmions profile when they are displacing under the action of a current density \cite{Troncoso}. Such a deformation can be represented by $\zeta-\xi\approx\delta$ (See Fig. \ref{Sk-deformados}). In this case, the mass matrix elements are given by $\mathcal{M}^{12}_{d}=\mathcal{M}^{21}_{d}=0$ and $\mathcal{M}^{11}_{d}=\mathcal{M}^{22}_{d}\equiv\mathcal{M}_d$. Assuming that $|\delta|\ll 1$, we can expand the mass elements of the $II$-skyrmions, neglecting terms of the order of $\delta^2$. Under these assumptions, we obtain that the mass elements of the $v$-bimeron configuration are
\begin{figure}
\includegraphics[scale=0.32]{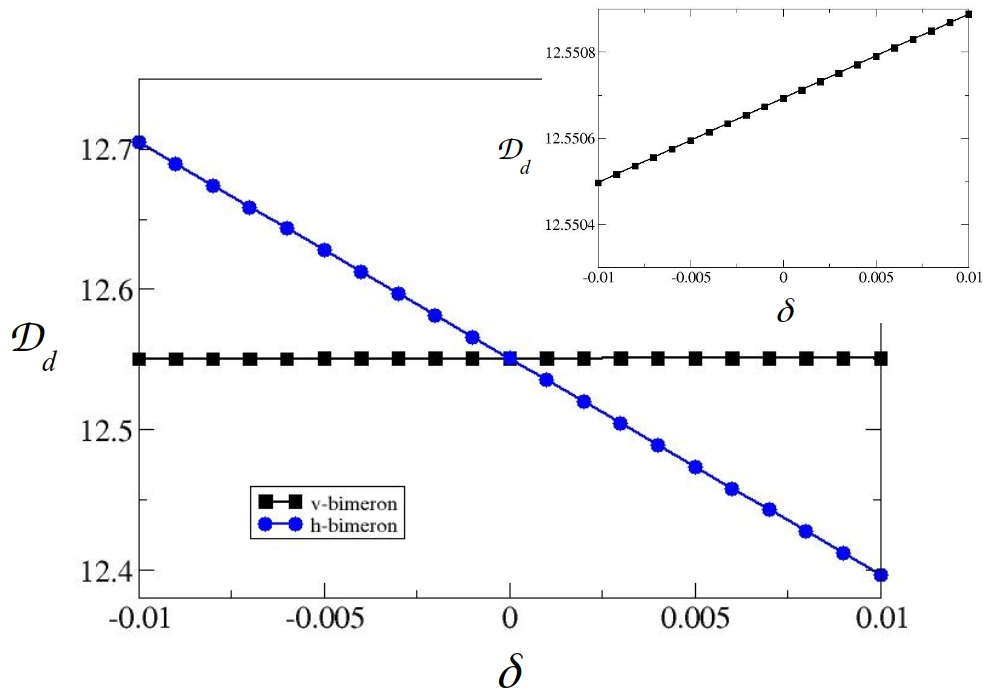}\caption{Behavior of $\mathcal{D}_d$ as a function of $\delta$. Black line (squares) represents the mass element of the $v$-bimeron. Blue line (circles) depicts the mass of a $h$-bimeron. The inset evidences that there is a variation in the $v$-meron mass. In the above figures, we have considered $R=4$ nm and $b=40$ nm.}\label{hmeron-delta}
\end{figure}

\begin{equation} \label{m11-d}
\mathcal{M}_d=\mathcal{M}_s+\frac{4\pi\delta\,b}{R^2}\left[2\sqrt{\mathcal{A}}\left(\frac{1}{\mathcal{A}}-2b^2+8\mathcal{A}b^4\right)-\frac{1}{\sqrt{\mathcal{B}}}-\frac{1}{\sqrt{\mathcal{C}}}\right]\,,
\end{equation}

\noindent
where $\mathcal{A}=(R^2+4b^2)^{-1}$, $\mathcal{B}=(R-2b)^{-2}$, and $\mathcal{C}=(R+2b)^{-2}$. The previous equation reveals that if the $v$-bimeron is flatten along the $x$-axis direction ($\zeta>\xi$), its mass increases, while if the $v$-meron is flattened along the $y$-axis direction ($\zeta<\xi$), its mass decreases. The mass elements for the $h$-bimeron can be also obtained. However, the equations describing them are cumbersome and  will be omitted here. In Fig. \ref{hmeron-delta} we show the behavior of $\mathcal{M}_d$ of the $h$-bimeron as a function of $\delta$. It can be observed that the mass elements of the $h$-bimeron behave contrary to the $v$-bimeron case. That is, for $\delta<0$ the mass increases when compared to the $\mathcal{M}_s$ and for $\delta>0$, the mass decreases. Additionally, the effect of the deformation on the mass is more prominent for $h$-bimerons.

From the above discussion, we are now in a position to explain the results obtained from micromagnetic simulations. Indeed, from the mass-center trajectory equation $y(x)= (\mathcal{D}_{s} \alpha /g) x \propto \mathcal{M}_s x$ (or Eqs. (\ref{xt})), one can observe that the position of the skyrmion depends on its mass in such a way that the larger the mass, more quickly the skyrmion approaches the lateral border of the racetrack. In principle, the annihilation of the structure at the racetrack lateral border would occur at a smaller $x$-position. In this context, because the $v$-bimeron mass practically does not change when it deforms, its trajectory should be almost the same as that of the type $I$-skyrmion. On the other hand, the $h$-bimeron diminishes its mass when it is flattened along the $y$-axis direction. Nevertheless, because the changes in the skyrmion mass are more pronounced for $h$-bimerons, the trajectory of this structure must have a more pronounced difference as compared with the type $I$-skyrmion pathway. Such results agree with the simulations. However, when all structures are near  the stripe border ($x \sim x_{i,c}$), the deformation along $x$-axis direction increases the $h$-bimeron mass and it is rapidly destroyed in the stripe border. Because we have considered a model for very small $\delta$, the trajectories obtained analytically are almost superposed, and then a most complete model should consider larger deformations. In addition, at $x \sim x_{i,c}$, the skyrmion-border interaction must also be very important for the skyrmion deformations, changing  drastically the skyrmion trajectories as indicated by the simulations.

\section*{Discussion and conclusion}

In summary, we have investigated how different skyrmion configurations travel along isotropic ferromagnetic racetracks. Since these skyrmions reside, in general, on different circumstances or materials (for instance, in-plane or out-of-plane boundary conditions dictate their structures), we have considered a race competition among them in which each skyrmion moves in its own appropriated lane. Since all the objects analyzed here experience the Hall skyrmion effect, they inevitably will die after running some distance along the racetrack (striking the lateral border). We show that the trajectories of these skyrmions depend on their mass, in such a way that, small modifications in the mass may result in an additional last breath, making determined skyrmion to live a bit more in the track. Our results show that a bimeron positioned in the $v$-bimeron mode is the best long-distance runner since it could go through a little more spatial extension before its annihilation at the lateral border of its racetrack. In spite the skyrmion-border interaction is not included, the presented theory gives a useful tool to understand the behavior of these different magnetic textures.

\section*{Acknowledgements}

The authors would like to thank (CAPES) - Finance Code 001, CNPq, FAPEMIG and also
the financial support from Financiamiento Basal AFB 180001 para Centros 
Cient\'ificos y Tecnol\'ogicos de Excelencia

\end{document}